\documentstyle[12pt]{article}
\begin{document}
\begin{titlepage}

\rightline{Preprint ITEP-M, October 1994}
\rightline{Preprint  IFUNAM FT 94-60 }
\rightline{hep-th/9410128}
\begin{center}
{\large Invariant identities in the Heisenberg algebra
\vskip 0.5cm
by
\vskip 0.5cm
 Alexander Turbiner}$^{\dagger }$
\vskip 0.5cm
 Institute for Theoretical and Experimental Physics, Moscow 117259, Russia\\
and\\
Instituto de Fisica, UNAM, Apartado Postal 20-364, 01000 Mexico D.F., Mexico
\end{center}

\begin{center}
{\large ABSTRACT}
\end{center}
\vskip 0.5 cm
\begin{quote}
Polynomial relations between the generators of $q$--deformed Heisenberg
algebra  invariant under the quantization and $q$-deformation are discovered.
\end{quote}

\vfill

\noindent
$^\dagger$E-mail: turbiner@teorica0.ifisicacu.unam.mx
or turbiner@vxcern.cern.ch

\end{titlepage}
Let in some algebra with the identity operator over the field of complex
numbers
there exist two elements $a$ and $b$, obeying the relation
\begin{equation}
\label{e1}
ab - q ba \ =\ p,
\end{equation}
where $p, q$ are any complex numbers. The algebra with the identity operator
generated by the elements $a,b$ and obeying (1) is usually named $q$--deformed
Heisenberg algebra $h_q$ ( see e.g. \cite{1}, references and a discussion
therein).
The parameter $q$ is named the parameter of quantum deformation. If $p=0$,
then (1) describes the non-commutative (quantum) plane (see e.g. \cite{2}),
while
if $q=1$ , then $h_q$ becomes the ordinary (classical) Heisenberg algebra (see
e.g.
\cite{3}) and the parameter $p$ plays a role of the Planck constant.
Then  the following theorem holds

{\bf \it THEOREM 1} \cite{4}.
For any $p,q \in {\bf C}$ in (\ref{e1}) the following identities take place
\renewcommand{\theequation}{2.{\arabic{equation}}}
\setcounter{equation}{0}
\begin{equation}
\label{e2.1}
(aba)^n\ = \ a^n b^n a^n , \quad n=1,2,3,\ldots
\end{equation}
and, if $q \neq0$, also
\begin{equation}
\label{e2.2}
(bab)^n\ = \ b^n a^n b^n , \quad n=1,2,3,\ldots
\end{equation}
\renewcommand{\theequation}{\arabic{equation}}
\setcounter{equation}{2}
The proof can be done  by induction by means of the following

{\bf \it  LEMMA 1}.
For any $p,q \in {\bf C}$ in (\ref{e1})  it holds
\renewcommand{\theequation}{3.{\arabic{equation}}}
\setcounter{equation}{0}
\begin{equation}
\label{e3.1}
a b^n \  -\ q^n b^n a \ = \ p \{ n \} b^{n-1}
\end{equation}
\begin{equation}
\label{e3.2}
a^n b \  -\ q^n b a^n \ =\  p \{ n \} a^{n-1}
\end{equation}
\renewcommand{\theequation}{\arabic{equation}}
\setcounter{equation}{3}
where $n=1,2,3 \ldots $ and $\{ n \}= {1-q^n \over 1-q}$ is the so-called
$q$-number
(see e.g. \cite{5}).

Now let us proceed to the proof of (2.1). At $n=1$ the relation (2.1) is
fulfilled trivially.
Assume that (2.1) holds for some $n$ and check the validity of (2.1) at $n+1$.
It is easy to write the chain of equalities :
\[
(aba)^{n+1}\ =\ (aba)^n aba \ =\ a^n b^n a^n aba\
=\ a^n (b^na)(a^nb)a\ =
\]
\[
=\ a^n ({ab^n \over q^n} - {p \over q^n} \{ n \} b^{n-1})
(q^n ba^n + p \{ n \} a^{n-1})a \ =\ a^{n+1} b^{n+1} a^{n+1}
\qquad (\diamondsuit)
\]
where the relations (3.1--2) were used in this chain. In analogous manner, one
can prove
(2.2). {\it q.e.d.}

The Theorem 1 leads to two corollaries. Both of them can be easily verified:

{\bf \it Corollary 1}.
For any $p,q \in {\bf C}$ in (\ref{e1})  and natural $n,k$ \ , it holds
\begin{equation}
\label{e4}
(\underbrace{ ababa....aba}_{2k+1})^n =
\underbrace{a^n b^n a^n ...b^n a^n}_{2k+1}  , \quad n,k=1,2,3,\ldots
\end{equation}

{\bf \it Corollary 2}.
Let  $T^{(n)}_k= \underbrace{a^n b^n a^n \ldots b^n a^n}_{2k+1} $,
then for any $p,q \in {\bf C}$ in (\ref{e1}),  it holds the relation
\renewcommand{\theequation}{5.{\arabic{equation}}}
\setcounter{equation}{0}
\begin{equation}
\label{e5.1}
 [ T^{(n)}_k , T^{(m)}_k ]\ = \ 0 \ ,
\end{equation}
and also more general relation
\begin{equation}
\label{e5.2}
  [ T^{(n_1)}_kT^{(n_2)}_k \ldots T^{(n_i)}_k ,
   T^{(m_1)}_k T^{(m_2)}_k\ldots T^{(m_j)}_k]\ = \ 0 \ ,
\end{equation}
\noindent
\renewcommand{\theequation}{\arabic{equation}}
\setcounter{equation}{5}
where $[\alpha, \beta] \equiv \alpha\beta - \beta \alpha$ is the standard
commutator
and  $<n>,<m>$ are sets of any non-negative, integer numbers.

It is evident that the formulas (4), (5.1-2) remain correct ones under
replacement
$a \rightleftharpoons  b$, if  $q\neq 0$.  It is worth noting that the algebra
$h_q$ under
appropriate choice of the parameters $p,q$ has a natural representation
\[
a = x\ ,\ b = D
\]
where the operator  $D f(x)={f(x)-f(qx) \over x(1-q)}$ is named the Jackson
symbol (see e.g.
\cite{5}). Then the relation (2.1) becomes
\[
(xDx)^n \ =\ x^n D^n x^n \ ,
\]
while the relation (2.2)
\[
(DxD)^n \ =\ D^n x^n D^n \ .
\]

Since  at $q \rightarrow1$,  the operator $D \rightarrow {d\over dx}$, hence
the above
relations become differential identities
\[
(x{d \over dx} x)^n \ =\  x^n {d^n \over dx^n} x^n
\]
and
\[
({d \over dx} x {d \over dx})^n \ =\ {d^n \over dx^n} x^n {d^n \over dx^n}
\]

{\bf \it THEOREM 2.}
For any $p,q \in {\bf C}$ in (\ref{e1})  and natural $n,m$  the following
identities  hold
\renewcommand{\theequation}{6.{\arabic{equation}}}
\setcounter{equation}{0}
\begin{equation}
\label{e6.1}
[a^nb^n, a^mb^m]\ =\ 0
\end{equation}
\begin{equation}
\label{e6.2}
[a^nb^n, b^ma^m]\ =\ 0
\end{equation}
\begin{equation}
\label{e6.3}
[b^na^n, b^ma^m]\ =\ 0
\end{equation}
\renewcommand{\theequation}{\arabic{equation}}
\setcounter{equation}{6}

The proof  is based on the following easy fact

{\bf \it LEMMA 2.}
For any $p,q \in {\bf C}$ in (\ref{e1})  and natural $n$
\renewcommand{\theequation}{7.{\arabic{equation}}}
\setcounter{equation}{0}
\begin{equation}
\label{e7.1}
a^n b^n \ =\ P(ab)
\end{equation}
\begin{equation}
\label{e7.2}
b^n a^n \ =\ Q(ab)
\end{equation}
\renewcommand{\theequation}{\arabic{equation}}
\setcounter{equation}{7}
where $P,Q$ are some polynomials in one variable of the order not higher than
$n$.

Let us introduce a notation $t^{(n)}=a^n b^n $ or  $b^n a^n$, so the order of
the
multipliers is not essential. Hence the statement of the Theorem 2 can be
written
as $[t^{(n)},t^{(m)}]=0$ and the following corollary holds

{\bf \it Corollary 3}.
The commutator
\begin{equation}
\label{e8}
 [ t^{(n_1)} t^{(n_2)} \ldots t^{(n_i)}, \  t^{(m_1)} t^{(m_2)} \ldots
t^{(m_j)}]\ = \ 0 \ ,
\end{equation}
for any  $p,q \in {\bf C}$ in (\ref{e1}) and any sets $<n>,<m>$  of
non-negative, integer
numbers.

One can make sense to (2.1--2), (4), (5.1--2), (6.1--3) as follows: in the
algebra of
polynomials in $a,b$ there exist relations invariant under a variation of the
parameters
$p,q$ in (1). Also the formulas (2.1--2), (4) can be interpreted as formulas of
a certain,
special ordering other than the standard, lexicographical one.

A natural question can be raised: an existence of the relations (2.1--2), (4),
(5.1--2),
(6.1--3), (8) is connected unambiguously to the algebra $h_q$, or there exists
more
general algebra(s) leading to those relations. The answer is given by the
following

{\bf \it THEOREM 3.}
If two elements $a,b$ of a certain Banach algebra with unit element are related
to
\begin{equation}
\label{e9}
ba = f(ab)
\end{equation}
where $f$ is a certain holomorphic function in a vicinity of variety
$\hbox{Spec}\,\{ab\} \cup \hbox{Spec}\,\{ab\} $, then the relations (2.1), (4),
(6.3) hold.
If an addition to that the function $f$ is single-sheeted one, then also (6.2)
holds.

The proof is based in essential on a fact, that once the function $f$ in (9) is
holomorphic,
then
\[
b F(ab)\ =\ F(ba) b
\]
what guarantees the correctness of the statement (7.2) of the Lemma 2, although
$Q$ is not polynomial anymore. It immediately proves (6.3). The relation (2.1)
can be
proven by induction and an analog the the logical chain ($\diamondsuit$) is
\[
(aba)^{n+1}\ =\ (aba)^n aba \ =\ a^n b^n a^n aba \ =\
a^n (b^na^n) aba\ =\
\]
\[
= \ [ b^na^n = Q (ab), {\hbox{see Lemma 2}}]\ =
\]
\[
 =\ a^n Q(ab) (ab) a \ =\ a^n (ab) Q(ab) a \ =\
a^{n+1} b^{n+1} a^{n+1}
\]
An extra condition that $f$ is single-sheeted implies that $ab = f^{-1}(ba)$.
It leads
immediately to the statement (6.2). {\it q.e.d.}

It is evident, that the replacement $a \rightleftharpoons b$ in the Theorem 3
leads to
the fulfillment of the equalities (2.2), (6.1) and then also (6.2),
correspondingly.

 In closing the author is very much indebted to R. Askey, A.M. Cetto, N.
Fleury,
A. Vershik and especially to N. Vasilevsky and S. L.~Woronovicz for useful
discussions
and for their interest in the subject. Also I am very grateful to the Instituto
de F\'isica,
UNAM for kind hospitality extended to me. This work was supported in part
by the research CONACyT grant.

\end{document}